\documentclass[prl,twocolumn,aps,superscriptaddress,showpacs]{revtex4-1}

\usepackage{amsmath,amssymb}
\usepackage{graphicx}
\usepackage{xcolor}
\usepackage{ulem}

\begin{document}

\title{\textbf{Metamagnetic Transition in UCoAl Probed by Thermoelectricity Measurements}}
\author{A. Palacio-Morales}
\email{alexandra.palaciomorales@cea.fr}
\affiliation{SPSMS, UMR-E CEA / UJF-Grenoble 1, INAC, Grenoble, F-38054, France}
\author{A. Pourret}
\affiliation{SPSMS, UMR-E CEA / UJF-Grenoble 1, INAC, Grenoble, F-38054, France}
\author{G. Knebel}
\affiliation{SPSMS, UMR-E CEA / UJF-Grenoble 1, INAC, Grenoble, F-38054, France}
\author{T. Combier}
\affiliation{SPSMS, UMR-E CEA / UJF-Grenoble 1, INAC, Grenoble, F-38054, France}
\author{D. Aoki }
\affiliation{SPSMS, UMR-E CEA / UJF-Grenoble 1, INAC, Grenoble, F-38054, France}
\affiliation{IMR, Tohoku University, Oarai, Ibaraki 311-1313,  Japan}
\author{H. Harima}
\affiliation{Department of Physics, Graduate School of Science, Kobe University, Kobe 657-8501, Japan}
\author{J. Flouquet}
\affiliation{SPSMS, UMR-E CEA / UJF-Grenoble 1, INAC, Grenoble, F-38054, France}

\date{\today }

\begin{abstract}
We report field and temperature dependent measurements of the thermoelectric power (TEP) and the Nernst effect in the itinerant metamagnet UCoAl. The magnetic field is applied along the easy magnetization $c$-axis in the hexagonal crystal structure. The metamagnetic transition from the paramagnetic phase at zero field to the field induced ferromagnetic (FM) state  is of first order at low temperatures and becomes a broad crossover above the critical temperature $T^{\star}_{M} \sim 11$~K. The field-dependence of the TEP reveals that the effective mass of the hole carriers changes significantly at the metamagnetic transition. The TEP experiment reflects the existence of different carrier types  in good agreement with band structure calculations and previous Hall effect experiments. According to the temperature dependence of the TEP, no Fermi liquid behavior appears in the paramagnetic state down to 150~mK, but is achieved only in the field induced ferromagnetic state.
\end{abstract}


\maketitle

The study of the quantum phase transitions (QPTs) in intermetallic strongly correlated electron systems such as the transition from an antiferromagnetic (AF) ordered state to a paramagnetic (PM), or from a ferromagnetic (FM) to PM ground states has motivated a large variety of experimental and theoretical studies recently \cite{Flouquet2005,Loehneysen2007,Si2010}. In the case of an AF instability, the restoration of a PM ground state at the QPT is often induced by application of pressure or magnetic field \cite{Stewart2006}. For FM materials at zero magnetic field the phase transition from PM to FM states changes from second order to first order at finite temperature before collapsing at the critical pressure $p_c$ \cite{Belitz1999}. Furthermore, the FM domain is extended above $p_c$ through the FM wings under external magnetic field \cite{Belitz2005, Taufour2010,Kotegawa2011}. These three FM first order planes define the location of the tricritical point at which the first order FM-PM transition ends up. A recent example is the FM compound $\textnormal{UGe}_2$ which at zero magnetic field has a tricritical point at $p_c\sim 1.46$~GPa \cite{Pfleiderer2002UGe2}. Under magnetic field the FM wings have been identified and they collapse at a quantum critical end point (QCEP) located around $p_\textnormal{QCEP} \sim 3.5$~GPa and $H \sim 16$~T \cite{Taufour2010, Kotegawa2011}.

Here we focus on the paramagnetic $5f$ system UCoAl which is located on the proximity of a FM quantum critical point, but the critical pressure from the FM to PM state is expected to be negative, $p_c\sim-0.2$~GPa \cite{Aoki2011UCoAl}. At a magnetic field $H_M \sim 0.7$~T  along the easy magnetization axis $c$ for $T\rightarrow 0$~K, a first-order metamagnetic transition from the PM to a FM ground state occurs \cite{Andreev1985}. The spontaneous field-induced magnetization is near $0.3 \mu_B/$U in the FM state \cite{Erisksson1989, Kucera2002}, much smaller than the effective moment from the Curie Weiss law above 30~K ($1.8 \mu_B/$U) indicating the itinerant character of the 5$f$ electrons \cite{Javorsky2001, Betsuyaku2000, Iwamoto2001}. The first order transition line $H_M (T)$ of this Ising type material terminates at a critical end point (CEP) with $H^{\star}_{M} \sim 1.0$~T and $T^{\star}_{M} \sim 11$~K. Above this CEP a crossover regime appears \cite{Aoki2011UCoAl}. It is worthwhile to notice that for $p<p_c$ the ground state of UCoAl may not be a pure FM state as it crystallizes in the hexagonal ZrNiAl-type structure (space group $P\bar{6}2m$) with lack of inversion symmetry \cite{Sechovsky1998}. In this structure the U ions are in a triangular coordination  forming a quasi-kagom\'{e} lattice within the $a$-$b$ plane. Thus an helical magnetic structure with a very short ordered wave vector, as e.g.~the one for MnSi \cite{Ishikawa1976, Stishov2011}, can be expected due to the Dzyaloshinsky-Moriya interaction.
		
In order to study in detail the electronic properties of UCoAl we performed precise thermoelectric power (TEP) experiments, extending previous measurements \cite{Matsuda2000UCoAl} from $T=4$~K down to $150$~mK. We carefully analyzed  the TEP in the different field  and temperature regimes of the $(T,H)$ phase diagram in order to draw the signature of $H_M(T)$ below the CEP ($H^{\star}_{M}$, $T^{\star}_{M}$) and of the PM-FM crossover  domain above the CEP. TEP measurements far above the Fermi liquid regime of the PM phase show that the ratio $S/T$ (TEP divided by temperature) has a relative stronger drop compared to the jump of the specific heat $C/T$ at $H_M$. Our results allow us to estimate the field variation of $S/T$ at very low temperatures  and thus to give a key comparison of the interplay between thermal transport and thermodynamic properties.

Single crystals of UCoAl were grown by the Czochralski method in a tetra-arc furnace.  The residual resistivity ratio of the studied crystals is around 10. Two bar-shaped samples from the same batch are cut by spark-cutter and oriented by X-ray Laue diffractometer displaying very sharp spots. TEP measurements were performed applying the magnetic field along the $c$-axis and thermal gradients along the $a$-axis (transverse case) and $c$-axis (longitudinal case).

The measurements are performed at low temperatures down to $150$~mK and under magnetic fields up to $5$~T. All data shown are obtained by sweeping the field upwards. Seebeck and Nernst coefficients, were measured using a ``one heater and two thermometers'' setup. Thermometers and heater are thermally decoupled from the sample holder by manganin wires.

\begin{figure}[t!]
\includegraphics[width=8cm]{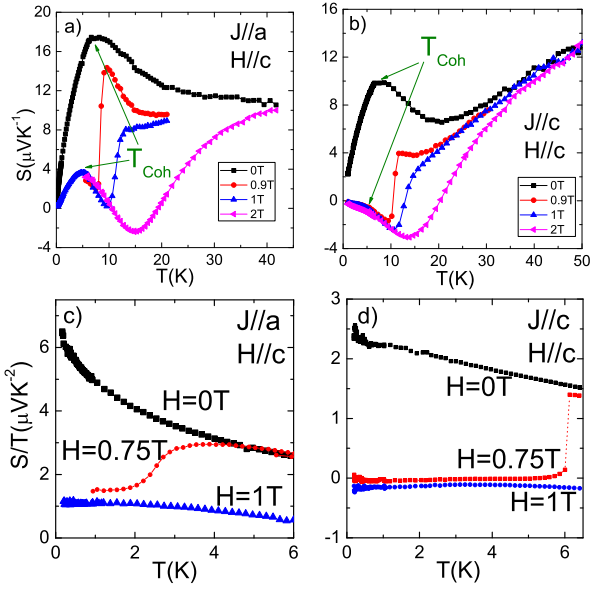}
\caption{\label{SvsT} (Color online). T dependence of the TEP $S(T)$ at different magnetic fields, a) for transverse and b) longitudinal thermal flow configurations. $S$ drops drastically at the metamagnetic transition and $T_{Coh}$ defined as the maximum of $S$, marks the entrance in the low temperature electronic regime where a Fermi liquid state is expected. c) and d) T dependence of the TEP divided by temperature  $S/T$, for transverse and longitudinal configuration, respectively. For both configurations, $S/T$ continues to increase down to 170~mK in the PM state for $H<H_{M}^{\star}(0)$ and gets constant above $H_{M}^{\star}(0)$ in the FM state.}
\end{figure}

\begin{figure}[t!]
	\includegraphics[width=8cm]{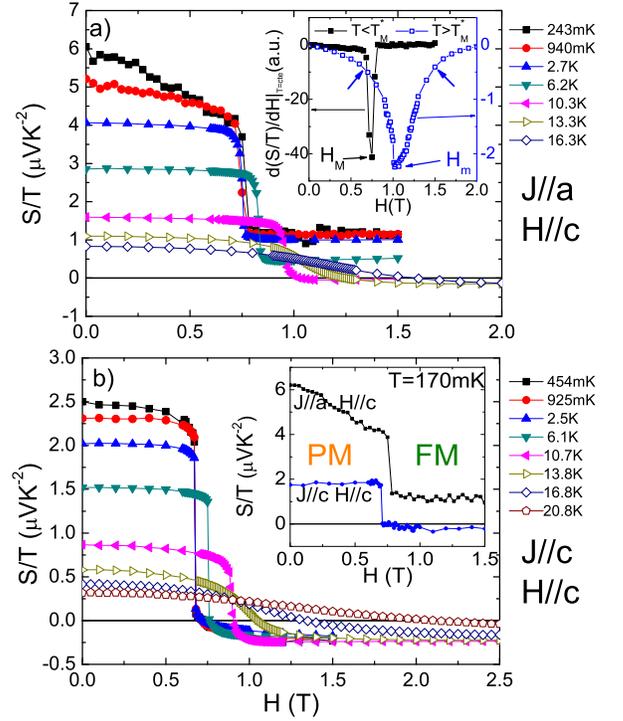}
	\caption{\label{SvsH} (Color online). Isothermal TEP measured as a function of increasing magnetic field at different temperatures in a) transverse and b) longitudinal configurations. The inset of panel~\ref{SvsH}a) indicates the location of the metamagnetic transition lines: $H_M$ ($1^{\rm st}$ order transition), $H_m$ (crossover); the two arrows delimit the broadness of the crossover. The inset of panel~\ref{SvsH}b) shows the field dependence of the TEP at the lowest temperature $\sim 170$~mK for $J\parallel a$ and $J\parallel c$ configurations.}
\end{figure}

Figures~\ref{SvsT}a) and b) show the temperature dependence of the TEP ($S$) at constant field for the transverse  and longitudinal configurations, respectively. The entrance in a low temperature domain where a Fermi liquid state is expected, is achieved below the temperature $T_{Coh}$ defined as the maximum of $S(T)$ in the transveral and a change of slope in the longitudinal configurations as indicated in Fig.~\ref{SvsT}a) and b).  Figures~\ref{SvsT}c) and d) show the temperature dependence of $S/T$ for $J\parallel a$ and $J\parallel c$ configurations, respectively. At $H=0$, a continuous increase of $S/T$ is observed down to the lowest temperature for both configurations. For the transverse configuration, the increase of $S/T$ is more pronounced and seems to diverge to low temperatures. 
 From the strong increase of $S/T$ on cooling at a constant field $H<H_M(0)$ it is clear that at least down to $150$~mK, a constant $S/T$ Fermi-liquid regime is not observed \cite{Zlatic2007}, neither for the longitudinal nor for the transverse configuration. By contrast for $H>H_M(0)$, a Fermi-liquid behavior is observed in the temperature dependence of $S/T$. 
The non-Fermi liquid behavior in the low field regime is also confirmed in resistivity measurements down to 50~mK (not shown). Previous resistivity experiments above $2$~K reported in \cite{Kolomiets1999} showed a $T^{3/2}$ dependence in the PM state while a Fermi liquid $T^2$ dependence occurs in the high field FM state. Similar behavior had been found in MnSi under high pressure. In that case, the resistivity has a $T^2$ dependence only in the ordered phase ($p < p_c$) while for $p>p_c \sim 1.4$~GPa a $T^{3/2}$ is observed over a large pressure range in the low field domain \cite{Doiron2003skyrmions, PedrazziniMnSINFL}. This unusual temperature dependence of the resistivity in MnSi has been attributed to a partial ordering observed in neutron scattering. It is interesting to notice for further pressure experiments on UCoAl that in MnSi up to $P \sim 3 p_c$, the $T^{3/2}$ dependence seems a robust pressure property of the low field ground state\cite{PedrazziniMnSINFL}. The very low temperature transport properties of UCoAl point at the formation of an exotic PM phase. This may be related to the peculiar quasi-kagom\'{e} lattice structure in this compound which can give rise to frustration as it has been stressed e.g.~for YbAgGe \cite{Sengupta2010}. 

Figure~\ref{SvsH} displays the isothermal TEP $S/T (H)$ as function of increasing field $H$ for the a) tranverse and b) longitudinal configurations. The inset in Fig.~\ref{SvsH} a) shows the determination of the 1$^{\rm st}$ order transition at $H_M$ and of $H_m$ (crossover) as well as the width of the crossover regime for $J\parallel a$ indicated by the arrows. The critical field of the samples used for the transversal configuration, $H_M = 0.75$~T, is slightly higher than that used for longitudinal configuration where we find $H_M = 0.7$~T. In both configurations, at the metamagnetic transition a strong hysteresis between up and down sweeps of the magnetic field due to the first order nature of the transition has been observed (not shown). The width of this hysteresis increases towards low temperatures (e.g.~40~mT at 0.85~K and 70~mT at 0.25~K for $J\;\parallel \;a$).
 
In order to estimate the discontinuity of $S(H)$ through the metamagnetic transition ($H_M$), we represent in the inset of Fig.~\ref{SvsH}b) the field dependence of $S/T$ at the lowest temperature achieved ($170$~mK) for both configurations. For $J\parallel c$, $S/T$ is almost field independent on both sides of $H_M$ while for $J\parallel a$ in the PM phase $S/T$ decreases with $H$ below $H_M$ and becomes constant above $H_M$. This different behavior of $S/T$ is clearly associated with the quite unusual strong increase of $S/T$ preserved at $H=0$~T for $J \parallel a$. For $J \parallel a$, the drop of $S/T$ as function of field at $T=170$~mK is at least $2.8$~$\mu$VK$^{-2}$ at $H_M$ while for $J \parallel c$ it is 2~$\mu$VK$^{-2}$. The step of the  specific heat coefficient ($\gamma = C/T$) is only $15\textnormal{mJ}\textnormal{mol}^{-1} \textnormal{K}^{-2}$ decreasing from $\gamma (H<H_M)\sim 75$~mJmol$^{-1}$K$^{-2}$ to $\gamma (H>H_M)\sim 60$~mJmol$^{-1}$K$^{-2}$ \cite{Aoki2011UCoAl, Matsuda2000UCoAl}. The relative drop in the TEP is $68$ $\%$ and  $100$ $\%$ for $J \parallel a$ and $J \parallel c$ configurations, respectively, while it is only $20$ $\%$ for $\gamma$. 
	
\begin{figure}[t!]
\includegraphics[width=8.5cm]{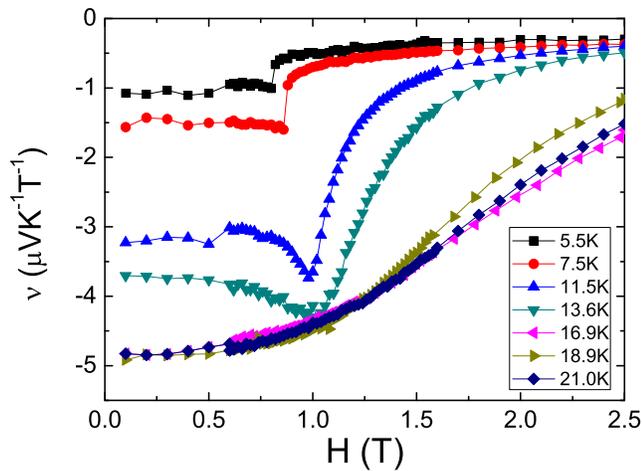}
\caption{(Color online). Isothermal Nernst coefficient, $\nu$ as a function of increasing field $H$. $\nu (H)$ presents an abrupt step at $H_M$ for $T<T^{\star}_{M}$ which becomes at $T>T^{\star}_{M}$ a pronounced minimum. This minimum broadens as the temperature increases in agreement with the increase of the broad crossover regime as shown in the phase diagram, see Fig.~\ref{PhaseDiagram}.}\label{Nernst}
\end{figure}	
		
\begin{figure}[t!]
\includegraphics[width=8.5cm]{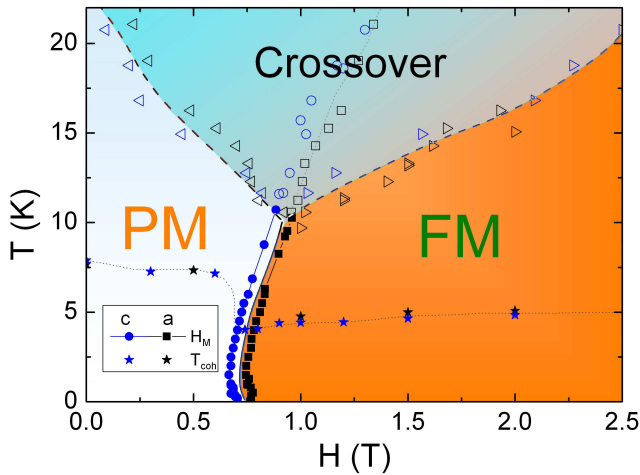}
	\caption{(Color online) $(T,H)$ phase diagram drawn for $J\parallel a$ (black points) and $J\parallel c$ (blue points) configurations; the observed critical field is slightly sample dependent. 
Open circles and open Squares are the cross-over field $H_m$. The grey dashed line and triangles indicate the crossover and the blue dotted marks the entrance in a coherent low temperature regime.  A simple Fermi liquid response ($S/T(T)=const$) is observed only for $H>H^{\star}_{M}(0)$ below $T_{coh}$.}\label{PhaseDiagram}
\end{figure}

 In Fig.~\ref{Nernst}, the Nernst coefficient $\nu=N/H$  is plotted as a function of magnetic field for temperatures around $T^{\star}_{M}$. The Nernst coefficient is almost constant at low magnetic fields which is consistent with the normal response of the Nernst effect in a PM state under magnetic field. At the metamagnetic transition for $T<T^{\star}_{M}$, the Nernst coefficient presents a sharp increase attributed to the presence of an anomalous Nernst effect (ANE)\cite{Onose2007, Lee2007, Onoda2008}. The transition becomes a pronounce  minimum for $T>T^{\star}_{M}$ which broadens as the temperature increases. The broadening range is consistent with the phase diagram determined by the TEP measurements (see Fig.~\ref{PhaseDiagram}). The observation of an anomalous Hall effect (AHE) is well known in ferromagnets \cite{Nagaosa2010}. It is the appearance of a spontaneous Hall current flowing parallel to \textbf{E} $\times$ \textbf{M}, where \textbf{E} is the electric field and \textbf{M} the magnetization. Karplus and Luttinger (KL) proposed that the AHE current is originated from an anomalous velocity term which is nonvanishing in a ferromagnet \cite{Luttinger1954}. The topological nature of the KL theory has been of considerable interest recently \cite{Onoda2002}. Similarly, the Nernst signal is also sensitive to the anomalous velocity term generating a dissipationless thermoelectric current, i.e.~an anomalous Nernst effect. The abrupt change of the Nernst signal in UCoAl at $H_{M}$ is similar to the anomaly observed in the Hall signal attributed to the AHE. Moreover the negative sign of the ANE is in good agreement with previous measurements of ANE e.g.~in the ferromagnet CuCr$_{2}$Se$_{4-x}$Br$_{x}$ \cite{Lee2004a}. 
 
The anomalies observed in the temperature and field dependence of the TEP are displayed in the $(T,H)$ phase diagram shown in Fig.~\ref{PhaseDiagram}. From the temperature dependent measurements at fixed field (see Fig.~\ref{SvsT}), we extract the cossover to the low temperature coherence regime.  From field dependence measurements (Fig.~\ref{SvsH}), the lines  $H_{M}$ and $H_{m}$ corresponding to the first order metamagnetic transition and the middle of the crossover, respectively. In addition the width of the crossover is shown in Fig.~\ref{PhaseDiagram}. 
Let us notice that, despite the complex $T$ dependence of $S(T)$ at fixed field, crossover lines can also be drawn from the $T$ dependences with, of course, slight differences from those determined from the field dependence of $S(H)$. However, the first order line coincides perfectly for the two determinations.
The crossover regime collapses on approaching the CEP ($H^{\star}_{M}$, $T^{\star}_{M}$). The entrance in a coherent regime appears at low temperatures and a simple Fermi-liquid state (constant $S/T(T)$ \cite{Zlatic2007}) occurs only for $H>H_M$. Above $1$~K, the phase diagram is in good agreement with those previously  drawn form magnetization, NMR, and Hall effect measurements \cite{Nohara2011,Karube2012,Combier2013}.  Below $T\sim 1$ K, a change of the sign in $\partial (T_M) / \partial H$ is due to the increase of the hysteresis of the first order transition.

The interest of TEP measurements is to probe the evolution of the topology of the Fermi surface, and the enhancement of the effective mass of the different types of carriers, electrons and holes. It is a rather difficult analysis in this multiband system. At first glance, it is usual to compare at very low temperature 
TEP and specific heat experiments. In a crude one-band model with a spherical Fermi surface, the TEP is directly linked to the entropy $S_e$ per charge carrier while the specific heat gives the entropy per mole. The ratio of both quantities defines the $q$-factor ($q=S N_A e/(\gamma T)$ where $N_A$ is Avogradro's number and $-e<0$ is the electronic charge) which is inversely proportional to the number of heat carriers per formula unit \cite{Behnia2004}. Taking the $\gamma$ value below and above $H_M$, for $H<H_M$ the $q$-factor is $q_a=0.12$ and $q_c=0.36$ for $J \parallel a$ and $J \parallel c$, respectively. Above $H_M$ $q_a$ jumps to $0.57$ and $q_c$ seems to change its sign becoming $-3.32$. These results suggest an important change through the metamagnetic transition and point out the irrelevance of a one-band description in this complex system. 

In order to improve previous full potential (FLAPW) band structure calculation \cite{Betsuyaku2000}, we have performed a more precise calculation by using 592 sampling k-points in the irreducible Brillouin Zone in the PM phase for this moderated heavy fermion compound. UCoAl is a compensated metal where the number of hole carriers, near $0.05$ holes per UCoAl formula, comes from the $78$ hole band  and the electron carriers are distributed among the electron bands $79$ and $80$. Although the LDA calculation cannot treat completely the electronic correlations, it will give a good estimation of the relative mass enhancement between hole and electron quasiparticles.
The density of states (DOS) of holes band is $260$ states/Ry/(primitive cell) corresponding to  15~mJmol$^{-1}$K$^{-2}$, while the DOS of the electron bands is $122$ states/Ry/(primitive cell) corresponding to 7~mJmol$^{-1}$K$^{-2}$, showing nearly twice heavier hole band than electron band. 

In a two band model with spherical Fermi surfaces the contribution of each band will be weighted by their respective electric conductivity ($\sigma $), i.e. $S=(\sigma_hS_h + \sigma_{e^{-}}S_{e^{-}})/(\sigma_h + \sigma_{e^{-}})$ \cite{Miyake2005}. 
In a first approximation, we assume the invariance of the Fermi surface through $H_M$. The hole carrier with an average effective mass $m_h^{\star} \approx 2m_e^{\star}$ will dominate the TEP response at low field in good agreement with the observed positive sign of the TEP.
Entering in the FM domain through $H_M$ will lead to a drastic decrease of $m_h^{\star}$, while the carrier concentration stays almost constant on crossing $H_M$
according to the Hall effect \cite{Matsuda2000UCoAl,Combier2013}. The opposite sign of the hole and electron response in $S/T$ leads to magnify the reduction of the TEP on entering in the FM domain and thus to a drop of $S/T$ quite stronger than the drop of $C/T$. 
However, a quantitative description is difficult as electron and hole Fermi surfaces are far to be spherical. Furthermore, in the FM polarized phase with a rather large FM component ($0.3 \mu _B$ per U atom), the spin up and spin down Fermi surfaces will differ. 

In future, an issue will be to test the validity of the Fermi surface invariance through $H_M$ via the direct observation of quantum oscillations or photoemission spectroscopy.
Recently, the interplay between Fermi surface topology and quantum singularities studied by TEP is also under debate in other heavy fermion systems (see e.g.~the metamagnetic transition in CeRu$_2$Si$_2$ \cite{Pfau2012}, the antiferromagnetic quantum critical point in YbRh$_2$Si$_2$ \cite{Hartmann2010}, or the hidden order in URu$_2$Si$_2$ \cite{Pourret2013}). 
 
In conclusion, TEP is a powerful probe to determine the ($T,H$) phase diagram of UCoAl down to very low temperature and far above the CEP at  $T_M^{\star}=11$~K and $H^{\star}_{M} \sim 1.0$~T. The metamagnetic transition is directly linked to the itinerant character of the quasiparticles. The comparison with Hall effect measurements below $H_M$ indicates that UCoAl is a multiband compound with a heavy hole band and a light electron band, in good agreement with band structure calculations. The main drop in the effective mass at the metamagnetic transition occurs  in the hole channel. A singular point is that due to its quasi-kagom\'{e} crystal structure with lack of inversion symmetry, the low field PM phase does not show Fermi-liquid properties even down to $150$~mK. The present data on the TEP in UCoAl with a rather simple Fermi surface may allow quantitative theoretical developments. It can serve as a reference for the recent TEP measurements made in strongly correlated electron systems. A clear route is now to realize measurements through the QCEP by tuning pressure and magnetic field.

We thank K.~Behnia for many useful discussions. We acknowledge the financial support of the French ANR (within the programs  PRINCESS, SINUS, CORMAT), the ERC starting grant (NewHeavyFermion), and the Universit\'{e} Grenoble-1 within the Pole SMINGUE.

\bibliographystyle{apsrev4-1}

\end{document}